# Graphene-enabled adaptive infrared textiles

M. Said Ergoktas[1,2], Gokhan Bakan[1,2], Pietro Steiner[1,2], Cian Bartlam[1,2], Yury Malevich[1,2], Elif Ozden Yenigun[3], Guanliang He[1], Nazmul Karim[4], Pietro Cataldi[2], Mark Bissett[1,2,6], Ian A. Kinloch[1,2,6], Kostya S. Novoselov[2,5], Coskun Kocabas[1,2,6,*]

1. Department of Materials, University of Manchester, Manchester, M13 9PL, UK
2. National Graphene Institute, University of Manchester, Manchester, M13 9PL, UK
3. School of Design, Textiles, Royal College of Art, London, SW7 2EU, UK
4. Centre for Fine Printing Research, University of the West of England, Bristol, BS16 1QY, UK
5. Department of Physics and Astronomy, University of Manchester, Manchester, M13 9PL, UK
6. Henry Royce Institute for Advanced Materials, University of Manchester, Manchester, M13 9PL, UK

Corresponding author: coskun.kocabas@manchester.ac.uk

**Interactive clothing requires sensing and display functionalities to be embedded on textiles. Despite the significant progress of electronic textiles, the integration of optoelectronic materials on fabrics still remains as an outstanding challenge. Here, using the electro-optical tunability of graphene, we report adaptive optical textiles with electrically controlled reflectivity and emissivity covering the infrared and near-infrared wavelengths. We achieve electro-optical modulation by reversible intercalation of ions into graphene layers laminated on fabrics. We demonstrate a new class of infrared textile devices including display, yarn and stretchable devices using natural and synthetic textiles. To show the promise of our approach, we fabricated an active device directly onto a t-shirt which enables long-wavelength infrared communication *via* modulation of the thermal radiation from the human body. The results presented here, provide complementary technologies which could leverage the ubiquitous use of functional textiles.**

The rapid progress of consumer electronics underpins the development of wearable technologies which have already reached a market value worth more than $50bn this year.[1] Textiles are the natural host material for emerging wearable devices. The key manufacturing challenge for smart wearable textiles is the large-scale integration of electronic/optical materials into fibres, yarns, and fabrics. Early works on embedding responsive materials such as shape memory alloys, pH or humidity sensitive polymers form the foundations of smart textile research.[2] The integration of devices into fabrics presents a more challenging task as

conventional wafer-based microfabrication methods require adaptations into soft and nonplanar forms.[3] Alternatively, wearable technologies can be developed as stand-alone units such as wearable accessories or stick-on patches on a variety of flexible substrates.[4] In contrast, integrating technologies directly into/onto clothes necessitates working with traditional textile materials and methods, while taking advantage of the ubiquity of these materials and well-established manufacturing infrastructure. Small-scale integrated circuits,[5] thermal management,[6,7] therapeutic patches,[8] sensors,[9,10] energy harvesting/storage[11] and haptic feedback[12] are among the common smart textile applications.

Integration of optoelectronic devices on textiles deserves a special mention as it promises applications across the electromagnetic spectrum ranging from visible[13–15] to infrared[16–19] even further to wireless communication wavelengths.[20,21] Applications using infrared light are especially appealing as the human body is a near-ideal black-body infrared source. Emerging technologies such as thermal management and adaptive camouflage require new abilities to regulate the infrared radiation from the surface while keeping the body temperature constant. New textile manufacturing technologies are being explored to achieve such capabilities. For example, Hsu *et al.* demonstrated enhanced radiative cooling of a human body using nanoporous polyethylene fabric.[16] Recently, they have also demonstrated polyethylene textile varying emissivity between ~0.9 and ~0.3 when flipped inside out.[17] In contrast, a dynamic modulation was reported by Zhang *et al.* using textile consisting of cellulose bimorph fibres and carbon nanotubes.[18] The infrared transmission is modulated between ~15% to ~35% by changing the relative humidity from 5% to 90%. A dynamic modulation of the infrared emissivity between ~0.95 and ~0.75 was demonstrated by Mao *et al.* based on thermally-triggered insulator-to-metal phase transition of $VO_2$ powder integrated into cotton fabric.[19] The state-of-the-art can be significantly improved by enhancing the modulation range and, more importantly, conceiving a more convenient method for controlling the infrared emission. Specifically, electro-optical control of the infrared emissivity can provide dynamic, fast, and on demand modulation that will enable integration with other wearable technologies, such as sensors and integrated circuits. However, this aim could not be achieved so far due to the lack of electro-optical infrared materials and technical challenges associated with the integration schemes.

Graphene and other 2-dimensional materials provide new opportunities for functional textiles.[22] The examples in the literature, however, rely on the electrical conductivity of these atomically thin materials. We have been investigating graphene as an adaptive optical platform operating from the visible to microwave wavelengths.[23–25] In our recent work, we have shown that thermal radiation from multilayer graphene can be modulated electrically *via* intercalation of ions.[26] Here, we introduce an optical textile technology by merging the electro-optical tunability of chemical vapour deposition (CVD)-grown graphene with novel textile devices. We show real-time electrical control of the infrared radiation in the wavelength range of 0.7 – 25 μm and reconfigurable infrared patterns from the device surface. The materials and the integration scheme reported here are compatible with the state-of-the-art large-area textile processing and a variety of textile materials, including, but not limited to, cotton, polyester, non-woven synthetics, conductive textiles, and yarns. These serve as not just a mechanical support, but also electrical separator, electrode, and ionic medium. The potential impact of the functional infrared textiles is highlighted by two showcase applications: merging sensing and display capabilities on a multipixel textile device and communicating a message in the long wavelength infrared by modulating the radiation from the human body.

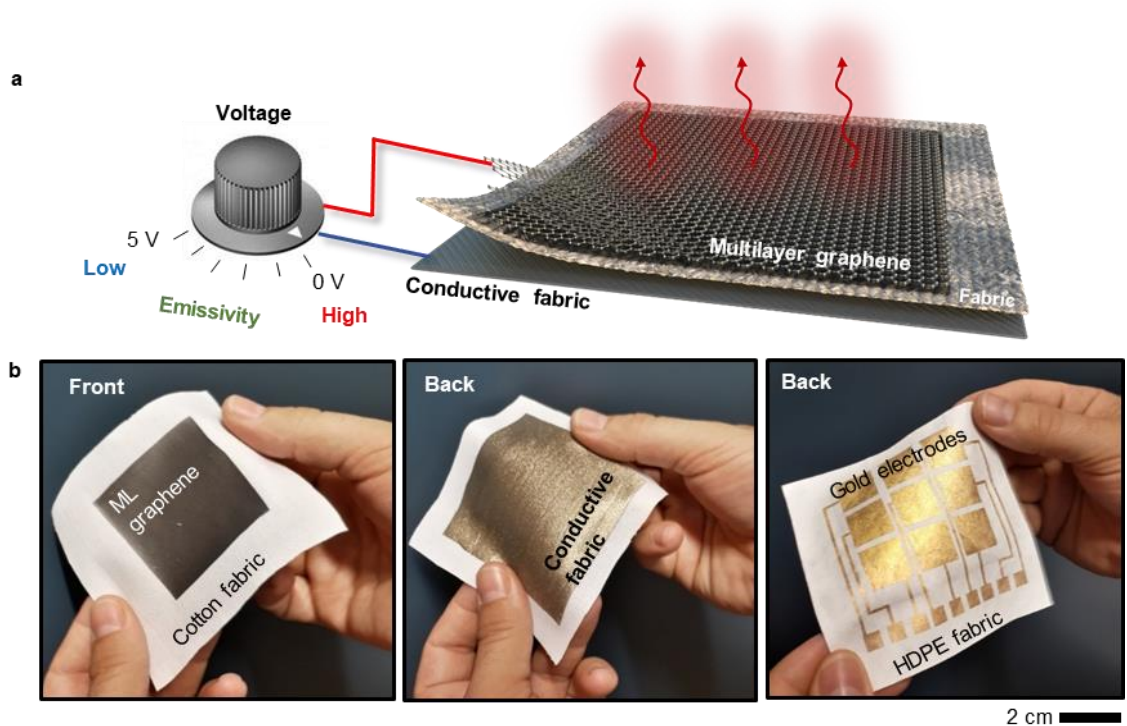

**Figure 1. Adaptive infrared textile**: **a**, Illustration of the textile device with various laminated layers: multilayer graphene, fabric separator and back electrode layer. The infrared transparent polymer coating on the graphene layer is not shown for clarity. The infrared emissivity of the device is modulated by applying a voltage difference between the top graphene layer and back electrode layer to initiate intercalation of ions into the graphene layers. **b,** Representative examples of fabricated devices on various textile materials such as woven cotton fabric and nonwoven high-density polyethylene fabric. Continuous conductive textiles or patterned gold electrodes can be used as the back electrode.

The devices consist of laminated layers of infrared transparent polymer layer, CVD-grown multilayer graphene (MLG), a fabric separator layer, and conductive fabric. Figure 1a shows a schematic drawing of a device and its operation. The fabrication starts with growing multilayer graphene layers on Ni foils (detailed in Materials and Methods section). A thin polyethylene (PE) film that functions as an infrared-transparent protective layer is laminated on multi-layer graphene prior to etching the Ni foil. MLG on polyethylene sheet is attached onto a fabric, e.g., a cotton twill weave, with hot melt adhesive. Good adhesion between MLG and the substrate is extremely important for any wearable application, thus we tested the stability of our MLG on textile under bending and during washing. To this end, some of the MLG on textile samples were subjected to mechanical bending/compression tests (Figure S1) and washing cycles (Figure S2, and Figure S3). The sheet resistance and infrared emissivity of MLG attached on cotton textile shows no sign of deterioration after the mechanical and washing tests, showing the high endurance and flexibility the devices. The back electrode, e.g., conductive fabric, is fused to the back side of the fabric forming the complete device (Figure 1a). In the final step, the ionic liquid electrolyte (BMIMPF$_6$) is applied onto the back electrode and allowed to diffuse into the textile substrate. The textile acts as both the separator and ionic conductive layer, allowing ionic motion when a voltage difference is applied to MLG and the back electrode. Figure 1b shows representative examples of the fabricated devices on the natural and synthetic textile materials, woven cotton and nonwoven high-density polyethylene clothing, respectively (see Figure S4 for more examples). We tested various back electrode materials including silver-based conductive textiles, stainless steel mesh, sputtered gold, graphene, and reduced graphene oxide. The electrochemical stability of the back electrode plays a critical role for the long term stability of the device. An array of patterned back electrodes and wiring on textile can be fabricated with photolithography followed by metallization and lift-off process (Figure 1b and Materials and Method section). These pixel electrodes allow us to define dynamic infrared patterns on a continuous graphene layer *via* area-selective intercalation.

The working principle of the devices is based on reversible intercalation of the ions into the graphene layers and modulating its electrical and optical properties (refer to Ref. 26 for an insight into the intercalation process). At 0 V, MLG has high infrared absorption which leads to high emissivity revealing the actual temperature of the device (Figure 2a). When a sufficient voltage difference is applied (> 2.5 V), the ionic liquid intercalates into the graphene layers, enhancing the optical conductivity and suppressing the emissivity, therefore concealing the actual device temperature. The thermographs of the device are recorded with a long wavelength infrared camera which renders images based on the Stefan–Boltzmann law, $P = \varepsilon\sigma T^4$, where $P$ is the amount of incident thermal radiation on the bolometer array, $\varepsilon$ is the emissivity of the surface kept constant during rendering, $\sigma$ is the Stefan–Boltzmann constant and $T$ is the temperature of the surface in degrees Kelvin. The temperature rendered by the camera (apparent temperature, $T_a$) is a function of $P$, hence $\varepsilon$ and $T$, and the emissivity setting of the camera, $\varepsilon_a$ (typically 1): $T_a = T(\varepsilon/\varepsilon_a)^{1/4}$. If the background emission reflected from the sample surface is non-negligible, it has to be accounted for. The textile devices are kept in thermal contact with the heat sources, e.g., human body, to prevent false screening of the source temperature. Furthermore, the MLG functions as a highly thermally conductive layer that doubles the in-plane thermal diffusivity of textile, enhancing heat conduction from the source to the surface (Figure S5). The temporal response of the devices was captured by recording thermal videos to obtain the variation in the apparent temperature of the surface (Figure 2b, Supplementary video 1). Complete intercalation (supressing emissivity) takes ~5 s when the device current is not limited. Note that these measurements were recorded in a closed lab environment which has background thermal radiation (21 °C) limiting the minimum apparent temperature (see Figure S6). The devices can be cycled steadily between high and low emissivity states many times (Figure 2c), however, exceeding the electrochemical window of the electrolyte degrades the device performance.

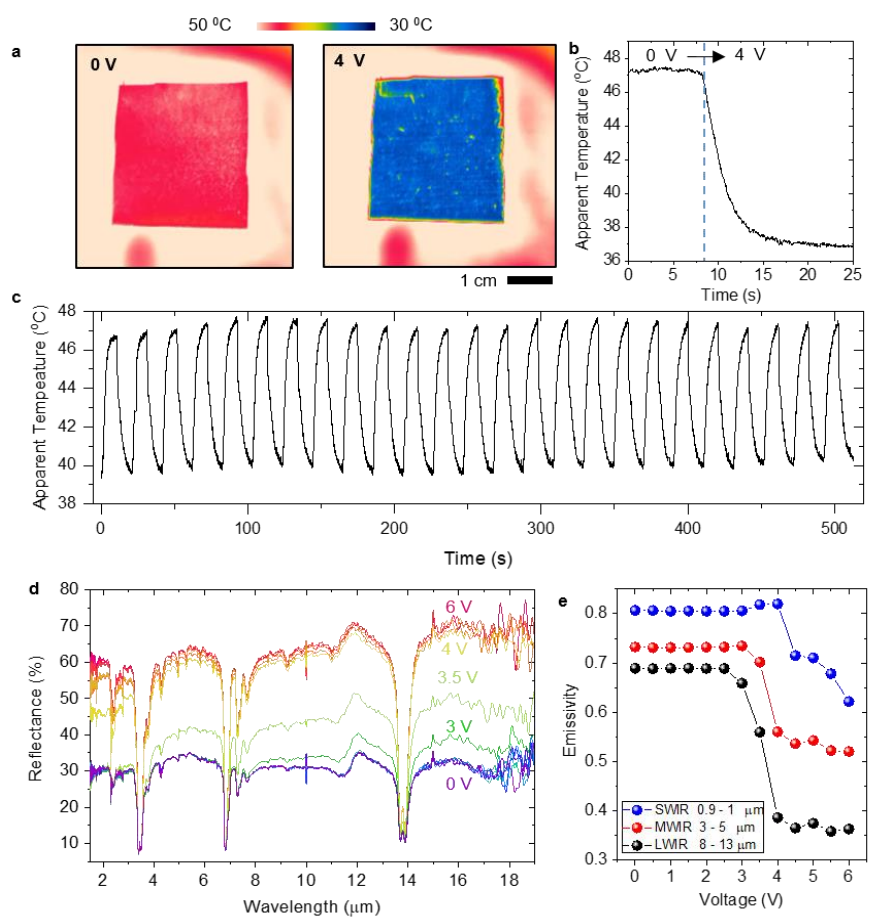

**Figure 2. Dynamic control of infrared emission on textile. a,** Thermal images of a cotton-based textile device recorded with a long-wavelength (8-13 μm) infrared camera in high emissivity (0 V) and low emissivity (4 V) states. The device is placed on a hot plate at 55 ℃. **b,** The temporal change in the apparent temperature of the device after applying 4 V. **c,** Long term temporal variation in the apparent temperature under a periodic square voltage waveform (-2 and 4 V) with a period of 20 s. **d,** Infrared reflection spectra of the device recorded as the applied voltage varies from 0 V to 6 V. **e,** The average emissivity of the device in the short, medium and long wavelength infrared regimes as a function of the applied voltage. The threshold voltage of the device is 2.5 V for LWIR.

The emissivity modulation was quantified by infrared and near-infrared reflection measurements using a Fourier-transform infrared spectrometer (FTIR) equipped with an integrating sphere. At 0 V, the reflectance of the intrinsic device is almost flat at ~30% except for the fingerprint absorptions of the top polyethylene film at wavelengths ~3.4 μm, ~6.8 μm, ~13.9 μm, and atmospheric absorptions, e.g. $CO_2$, $H_2O$ (Figure 2d). In the spectral sensitivity range of the thermal camera (8 – 13 μm), such absorptions are minimized owing to the careful selection of the top protection film (see Figure S7 for the transmission spectrum of the polyethylene film). The emissivity (or absorptivity) is calculated as $I−R$, where $R$ is the reflectance, as no light passes through the device. As the ions intercalate the graphene layers, the Fermi energy and the optical conductance of MLG increase, enhancing the infrared reflectance. The Pauli blocking of infrared absorption and the increased Drude optical conductivity of graphene are the main factors in the enhanced infrared reflectance.[27] The reflectance modulation is more pronounced for the longer wavelengths due to Drude type behaviour of free electrons on graphene. The average emissivity of the device in the wavelength range of 8 – 13 μm is high (~0.7) for 0 V and is maintained up to a threshold voltage (~2.5 V) followed by a sharp drop to ~0.35 for >4 V (Figure 2e). These results agree well with the thermographs shown in Figure 2a. The emissivity modulation covers both the long-wavelength infrared (LWIR, 8 – 13 μm) and medium-wavelength infrared (MWIR, 3 – 5 μm) ranges. In the MWIR, however, the polyethylene film exhibits a major absorption due to C–H stretching mode that is unaffected with the applied voltage, limiting the emissivity modulation range to ~0.7 – 0.5 (Figure 2e). Thus, applications in this wavelength range necessitate reconsideration of the protection layer. Another effect of the polyethylene layer is the enhanced emissivity of the surface owing to thermal extraction by polyethylene whose refractive index is larger than that of air.[28] We also observed relatively small emissivity modulation (0.2 – 0.4) in the short-wavelength infrared (SWIR, 0.9 – 1.7 μm) range. Nonetheless, the modulation in the SWIR can be detected by a silicon CCD camera with a near-infrared cut-on filter (Figure S8). The modulation in the visible spectrum is negligible due to insufficient doping of graphene. The optical modulation can be enhanced in the near-infrared and further pushed toward the visible regime with the use of a textile-compatible ionic liquid with a larger electrochemical window.

The electrically controlled emissivity of the textile devices together with the complex electrode patterns and embedded sensors can pave the way for interactive clothing. Such a device can serve multiple functions such as adaptive thermal camouflage or textile display. Figure 3 demonstrates such an example by combining the adaptive infrared response and display functions using a device with an array of 25 individually-addressable electrodes and a thermopile sensor (Figure 3a). A large single-piece MLG sheet on cotton fabric was used as the active layer (Figure 3b). Each electrode controls the emissivity of an area of $2 \times 2$ cm$^2$. An external electronic circuit was programmed to react to the heat signature on the sensor. The graphs in Figure 3c and 3d show the sensor signal and apparent temperature of an active pixel. The multipixel textile device shows letters "C" or "H" (representing cold and hot, respectively) by tuning the emissivity of corresponding pixels interacting to the absence or presence of a hot object over the sensor. Figure 3e shows the thermal images during the operation of the device as it alternates between the two letters when a hand is placed over the radiation sensor (see supplementary video 2). We also tested graphene-based capacitive touch sensors and strain sensors on textiles which could provide more complex feedback from the surrounding of the device. The demonstration highlights the capability of creating complex adaptive patterns and sensing capabilities which would inspire other applications such as adaptively blending the thermal signature of the device into a dynamic background.

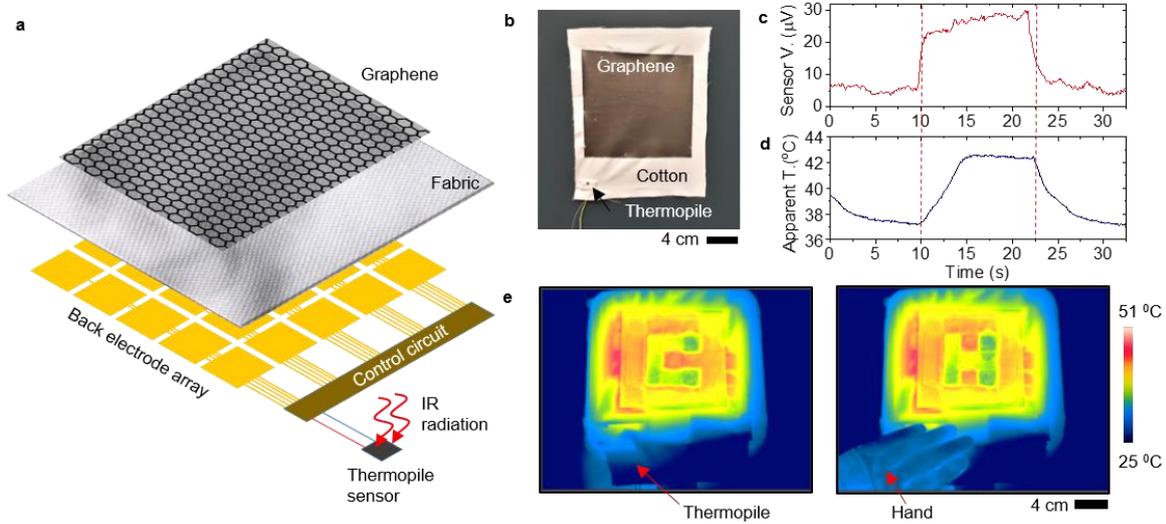

**Figure 3. Sensing and display on adaptive textile. a,** Illustration of a textile device with sensing and display capabilities. A large area graphene (15×15 cm²) layer was laminated onto woven cotton fabric. Individually-addressable 5×5 gold electrode array on the back of the textile was used to control the local emissivity of the device. A thermopile sensor was used to sense the presence of a hot object. An external electronic circuit processed the sensor signal and controlled the voltages of the individual pixels. **b,** Photograph of the device with the attached thermopile sensor. **c,** Variation of the output voltage of the sensor and **d,** the apparent temperature of an active pixel as a hot object, author`s hand, is brought over the sensor (t ≈ 10 s) and removed (t ≈ 22.5 s). The dashed lines mark the time when the sensor voltage reaches the pixel threshold voltage of 15 µV. **e,** Thermal images of the textile device displaying letters "C" and "H" as a response to the absence and presence of the hand, respectively. The temperature of the hot plate is 55 °C.

As our technology aims wearable smart textile devices, stretchability is crucial to accommodate for the natural deformation and drapeability of textile in mechanically active environments. Although graphene itself can sustain strain higher than 20%,[29] the CVD-grown polycrystalline multilayer graphene film is not stretchable due to the defects and grain boundaries. However, by structuring graphene layer into periodic, wavelike geometries on the textile surface can provide forms which can be stretched and compressed without damaging the graphene layer. We designed a stretchable textile device using a highly stretchable elastane fabric and a stretchable conducting fabric as a back electrode. MLG on PE sheets were laminated on the fully stretched elastane knitted fabric (82% polyester, 18% elastane) as described in Figure 4a. The MLG sheet is flat as fabricated and buckles when the fabric is relaxed, staying adhered to the fabric in both conditions (Figure 4b-c). In this wavelike buckled geometry, the device can support over 60% strain compared to the non-strained condition. The modulation of the apparent temperature on the complete device is demonstrated for relaxed and stretched conditions (Figure 4d). When the fabric is relaxed (zero strain) it exhibits higher emissivity (and absorptivity), hence higher apparent temperature, owing to enhanced light trapping by the buckled MLG (Figure 4e-f). The effect of buckling on the optical properties is also visible in Figure 4c.

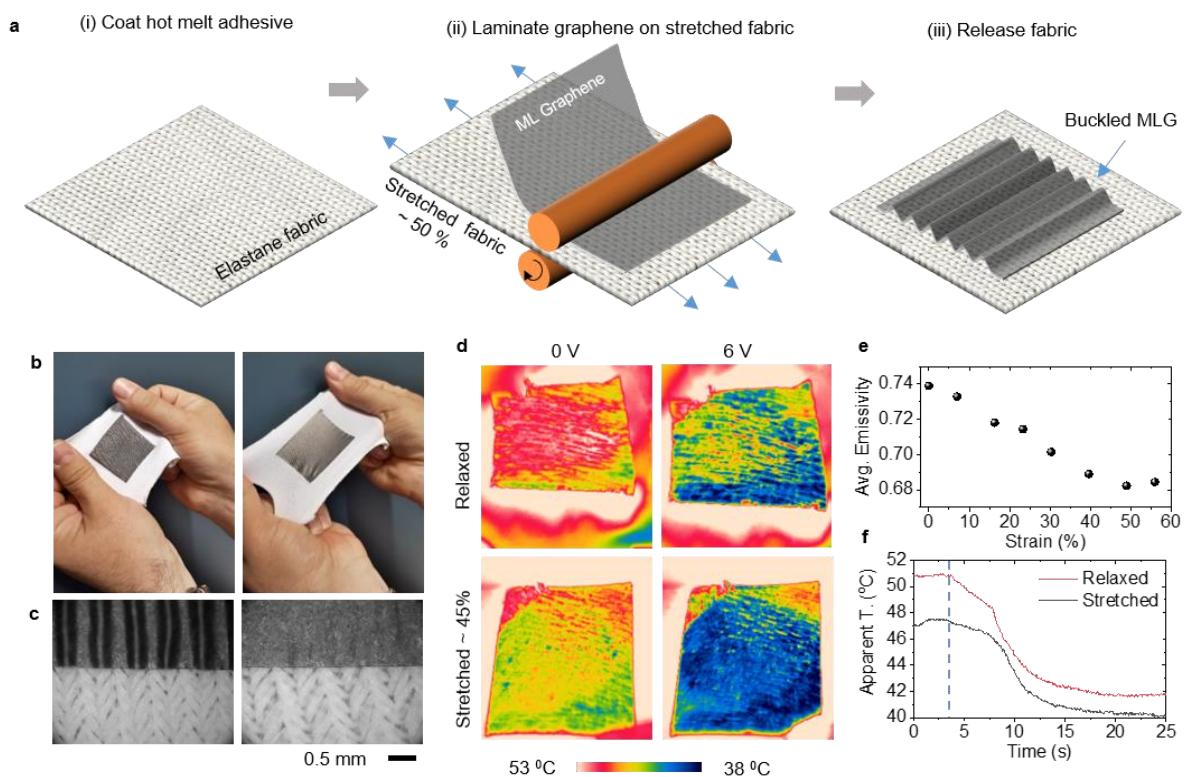

Figure 4. **Stretchable textile device**. **a,** Fabrication process: (i) a thin hot-melt adhesive layer was applied on the elastane fabric, (ii) MLG on PE sheet was laminated on the stretched fabric, (iii) after releasing the stress, the MLG/PE layer buckles and forms a wavy structure on the fabric. **b,** Photographs and **c,** microscope images of released and stretched device. The period of buckling is determined by the weaves of the fabric. **d,** Thermal images of the device showing electrically controlled thermal emission at the stretched and relaxed states. **e,** Variation of the emissivity (averaged between 8 – 13 μm) of the buckled graphene on elastane fabric as a function of strain. **f,** Temporal response of the apparent temperature when applying voltage to the device (marked by the vertical line) for relaxed and stretched conditions.

The simplicity of the reported fabrication process allows for it to be scaled down to yarn level, in turn enabling a finer spatial resolution and as well as forming an active textile surface by interlacing, e.g., knitting, weaving. The device structure necessitates yarns with conductive cores coated with textile materials. Amongst the options shown in Figure S9, yarns based on stainless steel wire and polyester cladding are chosen owing to the electrochemical stability of the stainless steel core and the uniformity of the polyester cladding. The conductive core and the cladding function as the back electrode and the separator/ionic-liquid-medium, respectively (Figure 5a). In this study a 270-μm-thick stainless steel wire is chosen for convenient handling of the yarns. Both the wire diameter and the cladding thickness can be scaled down for enhanced flexibility and ease of weaving. The yarns are covered with MLG using two methods: 1) Cutting MLG on PE sheets into narrow strips and winding them around the yarn, 2) Fishing MLG films by winding them around the yarns directly from the deionized water bath following the Ni-foil etching process (Figure 5b). The first method provides a better coverage of the yarn surface, however, this is laborious as it involves preparing the strips and winding them around the yarn carefully. The second method takes advantage of MLG films formed in the water bath following Ni-foil etching owing to the lack of a polymer protection layer. This method, however, results in an uneven surface coverage and more importantly an unprotected MLG layer that is prone to mechanical wear. The devices were finished by applying ionic liquid to the polyester cladding. The overall process is suitable for automation using motorized stages and is a necessity for large-scale production and precise control of winding MLG. The yarn devices necessitate a higher voltage difference (6 V) to turn on owing to the voltage drop over the long graphene layer. (Figure 5c, Figure 5d). Both MLG covering methods show similar performances in terms of the modulation speed and range. Different from the 2D textile devices demonstrated above, the yarn devices emit infrared radiation in all radial directions owing to their cylindrical shape. This feature can be an advantage for creating textile devices observable from any angle. However, if the application requires observing only one side of the device, as in the case of clothing, the fabrication scheme needs to be reconsidered for higher energy efficiency and operation speed.

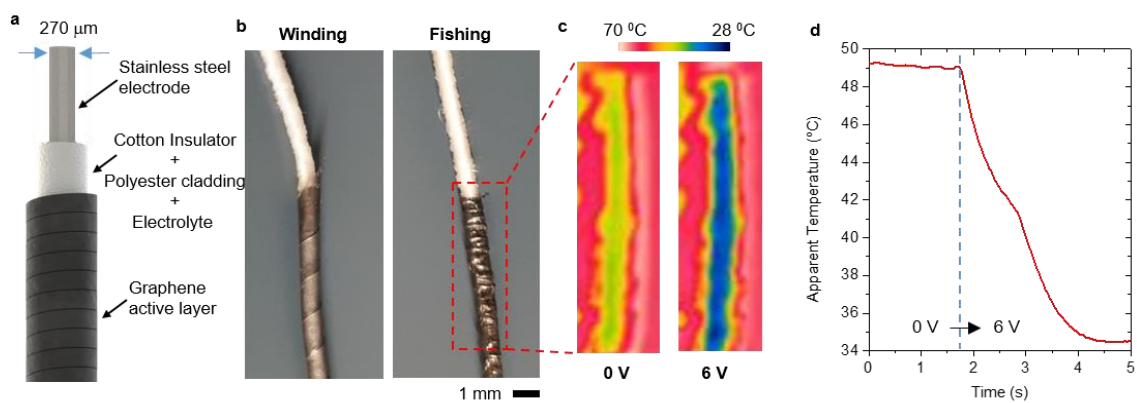

**Figure 5. Yarn devices. a,** 3D illustration of the yarn device consisting of a stainless-steel wire core, a porous polyester cladding with cotton insulation, and MLG outer layer. **b,** Photographs of two samples prepared by winding MLG/ PE strips and MLG films around the yarn: (i) MLG on PE substrate was cut into narrow strips and wound around the yarn, (ii) the yarn was covered with MLG films without PE substrate. **c,** Thermal images of a yarn device showing the modulated IR emission (Hot plate temperature is 70 ºC). **d,** The temporal change in the apparent temperature when turning it on (marked by the vertical line).

In the final part of the paper, to further highlight the promise of this adaptive textile, we demonstrate long wavelength infrared (LWIR) communication on a t-shirt by electrically modulating the infrared radiation of the human body. Due to natural body temperature, in ambient conditions, the human body radiates ~100 W of infrared light mainly in the LWIR range. This spectral range also coincides with the atmospheric transmission window which enables long distance propagation of the emitted LWIR light. Without using an additional light source, we were able to send messages undetectable by the naked eye or visible cameras by modulating the emissivity of a t-shirt. The t-shirt device was fabricated by laminating a $6 \times 6$ cm$^2$ MLG/PE film directly on a 100% cotton t-shirt surface and a stainless steel mesh to the back side (Figure 6a). We used a battery-powered microcontroller (Arduino UNO) to apply an encoded waveform generated by a pulse-width-modulated digital voltage that was programmed to communicate the initial letters of National Graphene Institute, "N", "G", and "I", in Morse code. The dash and dot signals were created by suppressing the apparent temperature for long (9 s) and short (3 s) durations. Figure 6b and 6c show the infrared snapshots of the t-shirt at the high and low emissivity states. The graph shows the recorded apparent temperature from a distance of 3 m. The experiment was conducted outdoors to take advantage of the lower background temperature.

This demonstration differs from the LWIR free-space optical communication that mainly aims transmitting data at high speed using high power infrared sources, e.g., quantum cascade lasers.[30] The main advantage of this demonstration is the use of the human body as a power source. Another advantage is that it prevents detection of the communicated message by the naked eye or visible cameras. The use of a microcontroller further allows building more sophisticated circuitry on textiles, in turn enabling more secure communication protocols, for instance initiation of the communication upon receiving an external triggering stimuli. The speed of communication using a single patch is limited by the intercalation/de-intercalation process which scales with the area of the device. Thus, the overall communication speed can be enhanced using multiple smaller patches and parallel processing of the message. Alternatively, a multipixel display, similar to the one employed in Figure 3, can be used to communicate alphanumerical characters or complex patterns. The size of a patch or a pixel depends on the imaging distance and the resolution of the thermal camera. This demonstration emphasizes the natural adaptability of the developed technology to everyday apparel such as a t-shirt in a realistic environment, the use of body temperature for the operation, and the portability of the technology through the use of a small-scale and light-weight controller and power source. Supplementary video 3 shows communication of the full message.

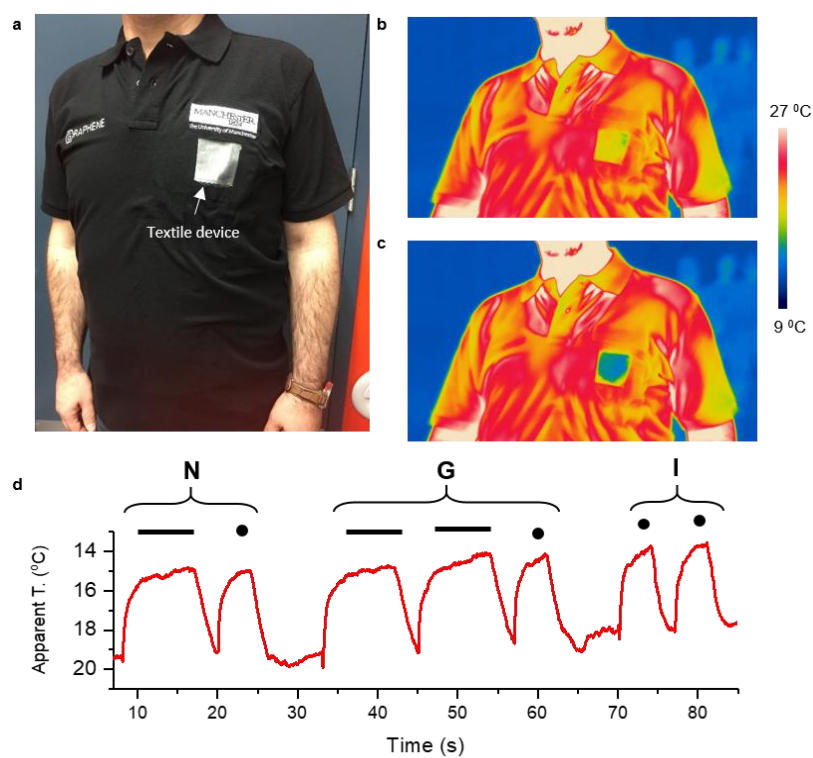

**Figure 6. Infrared communication on a t-shirt. a,** Photograph and thermographs for **b,** high emissivity and **c,** low emissivity states of a device fabricated on a t-shirt. The infrared emission from the author's body is modulated to communicate a message in the infrared. **d,** Modulation of the apparent temperature of the device communicating letters "N", "G", and "I" in Morse code. Temperature scale is reversed as the apparent temperature is suppressed when the device is on.

The use of small electrical signals for modulation of the infrared emissivity is a significant advantage over the alternatives as it enables an adaptive response, a necessity for dynamic thermal camouflage and thermal management applications. The developed technology requires low voltage (~3 V) and low energy ($5.5 \times 10^{-4}$ mAh/cm$^2$ per intercalation event, corresponding to a charge density of ~$10^{14}$ cm$^{-2}$ for each graphene layer, see Figure S10). Therefore, a compact coin cell with 1000 mAh capacity can turn on a t-shirt size (~1 m$^2$) device for ~180 times. Effectively operating as a capacitor, energy is consumed only during the charging (intercalation) cycle. Thus, the average standby power is virtually zero enabling hours or days of operation depending on the switching frequency. The low-voltage and low-power operation highly enhances the portability of the developed technology through the use of a miniature power supply and peripheral circuitry. In contrast, the alternative method (directly heating a surface) to control the thermal emission necessitates ~100 W for the same area. A coin cell battery can power such a device for only ~100 seconds. Directly cooling the surface with the help of Peltier coolers is even more energy consuming.[31]

In conclusion, this study provides a framework for graphene-enabled adaptive optical textiles. The developed methods for fabrication and transfer of graphene can easily be integrated into well-established textile infrastructure as the methods are compatible with the textile industry and large-area processing. The successful demonstration of the modulation of optical properties on different forms of textile can leverage the ubiquitous use of fibrous architectures and enable new technologies operating in the infrared and other regions of the electromagnetic spectrum for applications including communication, adaptive space suits, and fashion.

**Acknowledgements:** This research is supported by European Research Council through ERC-Consolidator Grant (grant no 682723, SmartGraphene). In addition, we acknowledge Graphene Engineering Innovation Centre (GEIC) for access to the CVD system.

**Author contributions**: C.K. conceived the idea. M.S.E. synthesized the graphene samples and fabricated the devices. M.S.E., G.B., and C.K. performed the experiments. G.B., M.S.E., K.S.N., and C.K. analysed the data and wrote the manuscript. P.S. conducted the thermal diffusivity measurements. E.O.Y. provided the textile samples and consultation on textile integration processes. G.H. and N.K. performed the mechanical tests. P.C. performed the washing tests. C.B., Y.M., M.B., and I.A.K. assisted with manuscript preparation. All authors discussed the results and contributed to the scientific interpretation as well as to the writing of the manuscript.

**Additional information**: C.K. is involved activities towards commercialization of variable emissivity surfaces under SmartIR Ltd. The work is subject to patent application by M.S.E., G.K., and C.K.. The remaining authors declare no competing financial interests.

## Materials and Methods

Growing MLG: MLG was synthesized on 25-μm-thick nickel foils (Alfa Aesar, 12722) by a chemical vapour deposition system (planarTECH CVD). First, a nickel substrate was heated to the growth temperature of 1050 °C under 100 sccm $H_2$, and 100 sccm Ar gases flow (quartz tube diameter 4″). Then, it was annealed at 1050 °C for 20 minutes to remove the native oxide layer. 35 sccm $CH_4$ flow at atmospheric pressure was used as the carbon precursor for 15 minutes. After the growth, the sample was cooled down to room temperature quickly under 100 sccm $H_2$ and 100 sccm Ar flow.

Transferring MLG on PE sheets: MLG on Ni foil was laminated at 160 °C on a 20-μm-thick polyethylene (PE) film that serves as a substrate for MLG during Ni foil etching and as well as an infrared-transparent protective layer once MLG was transferred on the fabric. Figure S11 illustrates the transfer process. To produce MLG on PE sheet, the Ni foil was etched in 1M $FeCl_3$ solution in ~8 hours. The uniformity of MLG films transferred on PE and on fabric were characterized using high-resolution infrared imaging (Figure S12). The apparent temperature distribution on each sample was used as a metric to quantify the uniformity of the infrared emissivity. MLG film on PE exhibits high uniformity similar to the performance of the black aluminium reference sample.

Fabrication of cotton-textile-based devices: The transfer onto the cotton fabric was performed by applying an adhesive layer on the fabric and laminating the MLG on PE sheet on. The devices were completed by adhering a conductive fabric on the other side functioning as the back electrode. The conductive and the cotton fabrics were adhered together using a thin, fusible, iron-on interfacing material in between and applying heat to fuse the fabrics. Electrical wires were connected to the MLG and the conductive fabric for electrical biasing. Conductive fabric was silver plated knitted fabric (Technik- Tex P). The ionic liquid electrolyte used was $BMIMPF_6$ (1-Butyl-3-methylimidazolium hexafluorophosphate, Sigma Aldrich 70956).

<u>Fabrication of elastane-textile-based devices</u>: Above procedure was followed with one additional step, where the elastane fabric was fully stretched while laminating MLG on a PE sheet.

<u>Fabrication of yarn devices</u>: Stainless steel soft wires (AISI 305, 0.27 mm) accompanied by insulating 100% cotton sheath yarns, (Ne 40) were uniformly covered by monofilament polyester at the twisting speed of 3000 twist/min (Agteks, DirectCover 2S). MLG was wrapped around polyester-cladded stainless-steel wires using two different methods: 1) MLG on PE sheets are created as described above and cut into narrow strips. Then the strips were wound around the yarn after applying ionic liquid electrolyte $BMIMPF_6$ to the polyester cladding. It is important to avoid overlapping the strips while wrapping the yarn to prevent unsuccessful intercalation at the edges, and 2) the PE lamination was omitted. This led to MLG forming films upon rinsing in a deionized water bath following Ni-foil etching. MLG films were then directly fished from the water by winding them around the yarn. Ionic liquid was applied around the MLG films.

<u>Fabrication of electrode arrays on textile</u>: 40-µm-thick negative dry photoresist was coated on nonwoven high-density polyethylene textile by hot lamination. Electrode array patterns on a transparent plastic stencil were transferred to the photoresist using a large-area ultraviolet exposure unit delivering 40 mJ/cm$^2$. The photoresist was developed in $K_2CO_3$ solution (5% concentration) for 2 minutes. The samples were coated with 100 nm Au films in a sputtering chamber (sputtering current: 20 mA, deposition rate: 13 nm/min). Finally, the remaining photoresist was lifted off leaving the desired patterns on the textile.

<u>Measurements:</u> The material characterization of MLG was performed using Raman spectroscopy (532nm laser, 2s exposure and 3accumulation) prior to the transfer process (Figure S13). The voltage difference on a complete device was controlled by a source meter while monitoring the device current. The back electrode was electrically grounded while applying a positive voltage on MLG for intercalation of ions. De-intercalation was achieved by either applying 0 V or a slightly negative voltage (-1 V) to MLG. FTIR measurements were performed using a Perkin Elmer Spectrum 100 FTIR spectrometer equipped with an integrating sphere and a liquid nitrogen cooled mercury-cadmium-telluride detector at a spectral resolution of 4 cm$^{-1}$. The reflection (*R*) results were used to evaluate the emissivity spectra (*ε = 1-R*). Figure S14, Figure S15, and Figure S16 show infrared reflection (R), transmission (T), and emissivity (ε) spectra of common textile materials. Emissivity for these materials was calculated as *1-R-T*. Thermal images and videos were recorded by a FLIR thermal camera (FLIR T660). The near infrared measurements were performed using another integrating sphere and a visible/NIR spectrometer. The NIR images were taken using a CCD camera without a NIR cut-off filter. A long-pass optical filter (cut-on wavelength = 700 nm) was used to capture only the near-infrared response of the devices.

<u>Mechanical tests:</u> An MLG on PE sheet with dimensions of $10 \times 4$ cm$^2$ was transferred on a cotton fabric. The sheet resistance of MLG was continuously measured as the sample was repeatedly bent and compressed with a tensile tester to monitor the mechanical durability. The electrical resistance of MLG was recorded with a National Instrument 9219 data acquisition card (NI, American) and was used to track the mechanical quality of the sample.